\documentclass[preprint,superscriptaddress,nofootinbib]{revtex4}
  \usepackage{amssymb,amsmath,graphicx,subfigure,color}
  \usepackage[colorlinks]{hyperref}
  
  \allowdisplaybreaks[4]
  \begin{document}
  \title{Study of the ${\Upsilon}(1S)$ ${\to}$ $DP$ decays}
  \author{Yueling Yang}
  \affiliation{Institute of Particle and Nuclear Physics,
              Henan Normal University, Xinxiang 453007, China}
  \author{Mingfei Duan}
  \affiliation{Institute of Particle and Nuclear Physics,
              Henan Normal University, Xinxiang 453007, China}
  \author{Junliang Lu}
  \affiliation{Institute of Particle and Nuclear Physics,
              Henan Normal University, Xinxiang 453007, China}
  \author{Jinshu Huang}
  \affiliation{School of Physics and Electronic Engineering,
              Nanyang Normal University, Nanyang 473061, China}
  \author{Junfeng Sun}
  \affiliation{Institute of Particle and Nuclear Physics,
              Henan Normal University, Xinxiang 453007, China}

  \begin{abstract}
  Inspired by the potential prospects of high-luminosity
  dedicated colliders and the high enthusiasms
  in searching for new physics in the flavor sector at
  the intensity frontier, the ${\Upsilon}(1S)$ ${\to}$
  $D^{-}{\pi}^{+}$, $\overline{D}^{0}{\pi}^{0}$ and
  $D_{s}^{-}K^{+}$ weak decays are studied
  with the perturbative QCD approach.
  It is found within the standard model that the branching
  ratios for the concerned processes are tiny, about
  ${\cal O}(10^{-18})$, and far beyond the detective
  ability of current experiments unless there exists
  some significant enhancements from a novel interaction.
  \end{abstract}
  \maketitle

  Searching for possible new physics (NP) beyond the standard model
  (SM) of particles from precise measurements with huge data
  is a popular fashion nowadays for experimentalists
  and theorists.
  The $b$ quark weak decays are ideal places to explore NP effects,
  because at the intensity frontier, there will be more than
  $10^{14}$ $b\bar{b}$ pairs with 300 $ab^{-1}$ dataset
  at LHCb \cite{1808.08865,1709.10308} and about $5{\times}10^{10}$
  $b\bar{b}$ pairs with 50 $ab^{-1}$ dataset at Belle-II
  \cite{1709.10308,1808.10567} in the near future.
  The $b$ rare decays usually have tiny branching ratios within SM,
  and are often used to search for NP, because an obvious deviation
  from SM predictions might be a smoking gun of NP.
  Of course, the precondition is a comprehensive investigation
  into specific processes within SM.
  According to the future experimental prospects, in this paper,
  we will study the ${\Upsilon}(1S)$ ${\to}$ $DP$ decays (here $P$
  $=$ ${\pi}$ and $K$) within SM in order to offer a ready reference
  for future analysis.

\begin{figure}[ht]
  \includegraphics[width=0.25\textwidth,bb=205 545 370 640]{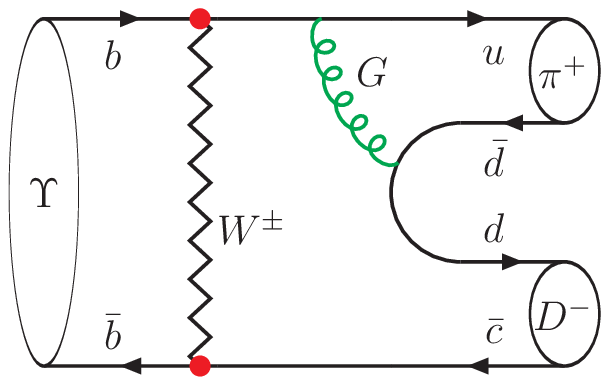}
  \caption{The Feynman diagram for the ${\Upsilon}(1S)$ ${\to}$
  $D^{-}{\pi}^{+}$ decay within SM.}
  \label{feynman-sm}
  \end{figure}

  The ${\Upsilon}(1S)$ meson is one of the $b\bar{b}$ bound states
  (bottomonium). The mass of ${\Upsilon}(1S)$ meson,
  $m_{{\Upsilon}(1S)}$ $=$ $9460.30(26)$ MeV \cite{pdg2020},
  is less than the open flavor threshold.
  The predominant decay is through the annihilation of the
  $b\bar{b}$ pairs into three gluons, with branching ratio
  ${\cal B}r({\Upsilon}{\to}ggg)$ $=$ $81.7(7)\%$
  \cite{pdg2020}.
  However, the hadronic decays are hindered by the phenomenological
  Okubo-Zweig-Iizuka (OZI) rule \cite{ozi-o,ozi-z,ozi-i},
  and result in the extremely narrow decay width,
  ${\Gamma}_{{\Upsilon}}$ $=$ $54.02(1.25)$ keV \cite{pdg2020}.
  To date, the sum of measured branching ratio of 100 exclusive
  hadronic modes is only about 1.2\% \cite{pdg2020}.
  Besides the strong and electromagnetic transitions,
  the ${\Upsilon}(1S)$ meson can decay through the weak
  interactions, for example, the flavor non-conservation
  processes of ${\Upsilon}(1S)$ ${\to}$ $DP$ decays.
  It is estimated that the branching ratios of ${\Upsilon}(1S)$
  weak decays should usually be very small, about
  $2/{\tau}_{B}{\Gamma}_{{\Upsilon}}$ ${\sim}$ ${\cal O}(10^{-8})$,
  where ${\tau}_{B}$ and ${\Gamma}_{{\Upsilon}}$
  are the lifetime of $B$ meson and width of ${\Upsilon}$ meson.
  Within SM, the ${\Upsilon}(1S)$ ${\to}$ $DP$ decays are
  induced by the $W^{\pm}$ exchange,
  which is illustrated in Fig. \ref{feynman-sm}.

  A sample of $10^{8}$ ${\Upsilon}(1S)$ events has been
  collected at resonances by the Belle detector \cite{epjc74.3026}.
  A much more number of ${\Upsilon}(1S)$ with great precision is
  expected at the running Belle-II and upgraded LHCb experiments.
  Besides the direct production via $e^{+}e^{-}$ ${\to}$
  ${\Upsilon}(1S)$, the ${\Upsilon}(1S)$ meson can also be
  produced via the ${\Upsilon}(nS)$ ${\to}$
  ${\pi}{\pi}{\Upsilon}(1S)$ and ${\eta}{\Upsilon}(1S)$
  transitions (where $n$ ${\ge}$ 2) and initial state radiation
  processes $e^{+}e^{-}$ ${\to}$ ${\pi}{\pi}{\Upsilon}(1S)$.
  The huge amount of data make the study of ${\Upsilon}(1S)$
  weak decay interesting and worthwhile, although very challenging.

  With the help of the excellent performance of Belle-II and LHCb
  detectors, and the assistance of sophisticated analysis
  technology and methods, events of the ${\Upsilon}(1S)$
  ${\to}$ $DP$ decays should in principle be easily selected.
  On the one hand, the final states carry definite energies
  and momenta in the rest frame of the ${\Upsilon}(1S)$ meson;
  on the other hand, the identification of a single charmed
  meson is free from inefficiently double tagging, and provides
  a conclusive evidence of the ${\Upsilon}(1S)$ weak decay.
  And what's more, the phenomenon of an abnormally large production
  rate of a single charmed mesons would be a hint of NP.

  As far as we know, the ${\Upsilon}(1S)$ ${\to}$ $DP$ decays
  have not been studied seriously yet.
  From the experimental point of view, inadequate data samples
  and tiny branching ratios might be the main considerations.
  From the theoretical point of view, one of the principal problems
  is how to properly calculate the hadron transition matrix elements
  due to our limited informations about the hadronization mechanisms,
  the long-distance contributions, and so on.

  From Fig. \ref{feynman-sm}, it is clearly seen that there are
  simultaneously many scales involved in the theoretical calculation
  of the ${\Upsilon}(1S)$ ${\to}$ $DP$ decays, such as the mass of
  $W$ gauge boson $m_{W}$, the $b$ quark mass $m_{b}$ and the
  QCD characteristic scale ${\Lambda}_{\rm QCD}$.
  In general, different dynamics correspond to different scales.
  Here, we will adopt the commonly acknowledged treatment by using
  the effective theory.
  The effective Hamiltonian in charge of the ${\Upsilon}(1S)$
  ${\to}$ $DP$ decays is written as \cite{rmp68.1125},
   \begin{equation}
  {\cal H}_{\rm eff}\ =\ \frac{G_{F}}{\sqrt{2}}\,
   V_{ub}\,V_{cb}^{\ast}\, \sum\limits_{i=1}^{10}\,
   f_{i}\,C_{i}({\mu})\,O_{i}({\mu}) + {\rm H.c.}
   \label{eq:hamilton},
   \end{equation}
  where $G_{F}$ ${\simeq}$ $1.166{\times}10^{-5}\,{\rm GeV}^{-2}$
  \cite{pdg2020} is the Fermi coupling constant.
  The Cabibbo-Kobayashi-Maskawa (CKM) factor
  ${\vert}V_{ub}\,V_{cb}^{\ast}{\vert}$ $=$ $1.463(54){\times}10^{-4}$
  \cite{pdg2020}.
  The factor $f_{i}$ $=$ $+1$ for tree operators $O_{1,2}$
  and $-1$ for penguin operators $O_{3-10}$, respectively.
  The Wilson coefficients $C_{i}$ are calculable with the
  renormalization group improved perturbation theory at the
  scale of $m_{W}$, and then evolved to the scale of ${\mu}$.
  The operators describing the local interactions among four
  quarks are defined to be,
   \begin{eqnarray}
   O_{1} &=&
   \big[\bar{u}_{\alpha}\,{\gamma}_{\mu}\,
    (1-{\gamma}_{5})\, b_{\alpha} \big]\,
   \big[ \bar{b}_{\beta}\,{\gamma}^{\mu}\,
    (1-{\gamma}_{5})\,  c_{\beta} \big]
   \label{operator-01}, \\
   O_{2} &=&
   \big[\bar{u}_{\alpha}\,{\gamma}_{\mu}\,
    (1-{\gamma}_{5})\, b_{\beta} \big]\,
   \big[ \bar{b}_{\beta}\,{\gamma}^{\mu}\,
    (1-{\gamma}_{5})\,  c_{\alpha} \big]
   \label{operator-02}, \\
   O_{3} &=&
   \big[\bar{u}_{\alpha}\,{\gamma}_{\mu}\,
    (1-{\gamma}_{5})\, c_{\alpha} \big]\,
   \sum\limits_{q}\,
   \big[ \bar{q}_{\beta}\,{\gamma}^{\mu}\,
    (1-{\gamma}_{5})\,  q_{\beta} \big]
   \label{operator-03}, \\
   O_{4} &=&
   \big[\bar{u}_{\alpha}\,{\gamma}_{\mu}\,
    (1-{\gamma}_{5})\, c_{\beta} \big]\,
   \sum\limits_{q}\,
   \big[ \bar{q}_{\beta}\,{\gamma}^{\mu}\,
    (1-{\gamma}_{5})\,  q_{\alpha} \big]
   \label{operator-04}, \\
   O_{5} &=&
   \big[\bar{u}_{\alpha}\,{\gamma}_{\mu}\,
    (1-{\gamma}_{5})\, c_{\alpha} \big]\,
   \sum\limits_{q}\,
   \big[ \bar{q}_{\beta}\,{\gamma}^{\mu}\,
    (1+{\gamma}_{5})\,  q_{\beta} \big]
   \label{operator-05}, \\
   O_{6} &=&
   \big[\bar{u}_{\alpha}\,{\gamma}_{\mu}\,
    (1-{\gamma}_{5})\, c_{\beta} \big]\,
   \sum\limits_{q}\,
   \big[ \bar{q}_{\beta}\,{\gamma}^{\mu}\,
    (1+{\gamma}_{5})\,  q_{\alpha} \big]
   \label{operator-06}, \\
   O_{7} &=&
   \big[\bar{u}_{\alpha}\,{\gamma}_{\mu}\,
    (1-{\gamma}_{5})\, c_{\alpha} \big]\,
   \sum\limits_{q}\,\frac{3}{2}\, Q_{q}\,
   \big[ \bar{q}_{\beta}\,{\gamma}^{\mu}\,
    (1+{\gamma}_{5})\,  q_{\beta} \big]
   \label{operator-07}, \\
   O_{8} &=&
   \big[\bar{u}_{\alpha}\,{\gamma}_{\mu}\,
    (1-{\gamma}_{5})\, c_{\beta} \big]\,
   \sum\limits_{q}\,\frac{3}{2}\, Q_{q}\,
   \big[ \bar{q}_{\beta}\,{\gamma}^{\mu}\,
    (1+{\gamma}_{5})\,  q_{\alpha} \big]
   \label{operator-08}, \\
   O_{9} &=&
   \big[\bar{u}_{\alpha}\,{\gamma}_{\mu}\,
    (1-{\gamma}_{5})\, c_{\alpha} \big]\,
   \sum\limits_{q}\,\frac{3}{2}\, Q_{q}\,
   \big[ \bar{q}_{\beta}\,{\gamma}^{\mu}\,
    (1-{\gamma}_{5})\,  q_{\beta} \big]
   \label{operator-09}, \\
   O_{10} &=&
   \big[\bar{u}_{\alpha}\,{\gamma}_{\mu}\,
    (1-{\gamma}_{5})\, c_{\beta} \big]\,
   \sum\limits_{q}\,\frac{3}{2}\, Q_{q}\,
   \big[ \bar{q}_{\beta}\,{\gamma}^{\mu}\,
    (1-{\gamma}_{5})\,  q_{\alpha} \big]
   \label{operator-10},
   \end{eqnarray}
  where ${\alpha}$ and ${\beta}$ are color indices and the
  sum over repeated indices is understood.
  $Q_{q}$ is the electric charge of quark $q$ in the unit
  of ${\vert}e{\vert}$, and $q$ ${\in}$
  \{$u$, $d$, $c$, $s$, $b$\}.

  With the interaction Hamiltonian of Eq.(\ref{eq:hamilton}),
  the decay amplitudes for the ${\Upsilon}(1S)$ ${\to}$ $DP$
  decays can be written as,
   \begin{equation}
  {\cal A}({\Upsilon}{\to}DP)\, =\,
  {\langle}DP{\vert} {\cal H}_{\rm eff}
  {\vert} {\Upsilon} {\rangle}\, =\,
   \frac{G_{F}}{\sqrt{2}}\,
   V_{ub} V_{cb}^{\ast}\, \sum\limits_{i=1}^{10}\,
   f_{i}\,C_{i}({\mu})\,{\langle}DP{\vert} O_{i}({\mu})\,
  {\vert} {\Upsilon} {\rangle}
   \label{eq:hamilton-amplitudes}.
   \end{equation}
  It is seen that the decay amplitudes of
  Eq.(\ref{eq:hamilton-amplitudes}) are clearly factorized into
  four parts: the couplings of weak interactions $G_{F}$,
  the CKM factors $V_{ub} V_{cb}^{\ast}$, the Wilson
  coefficients $C_{i}$ summarizing the physical contributions
  above the scale of ${\mu}$, and the hadronic matrix elements (HMEs)
  ${\langle}O_{i}{\rangle}$ $=$ ${\langle}DP{\vert} O_{i}({\mu})\,
  {\vert} {\Upsilon} {\rangle}$ containing the physical
  contributions below the scale of ${\mu}$.
  The product of the first three parts, $G_{F}$,
  $V_{ub}V_{cb}^{\ast}$ and $C_{i}$, can be regarded as
  the effective coupling of operators $O_{i}$,
  and has been well known.
  The HMEs ${\langle}O_{i}{\rangle}$ describing the transitions
  from quarks to participating hadrons are the core and difficulty
  of theoretical calculations.
  In addition, the QCD radiative corrections to HMEs should be
  included in order to obtain a physical amplitude by cancelling
  the scale ${\mu}$ dependence of the Wilson coefficients.

\begin{figure}[ht]
  \includegraphics[width=0.22\textwidth,bb=200 535 370 635]{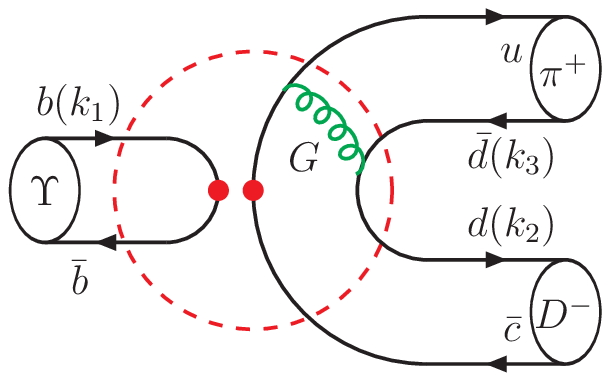}\quad
  \includegraphics[width=0.22\textwidth,bb=200 535 370 635]{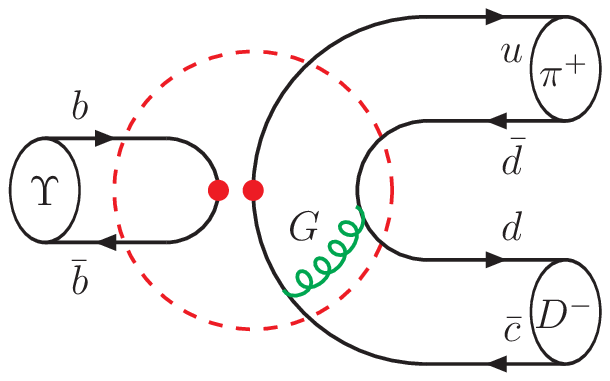}\quad
  \includegraphics[width=0.22\textwidth,bb=200 535 370 635]{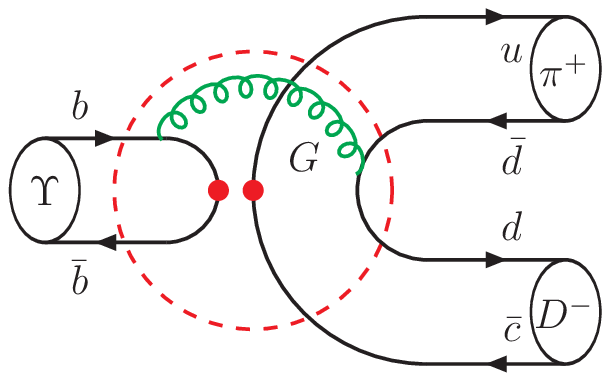}\quad
  \includegraphics[width=0.22\textwidth,bb=200 535 370 635]{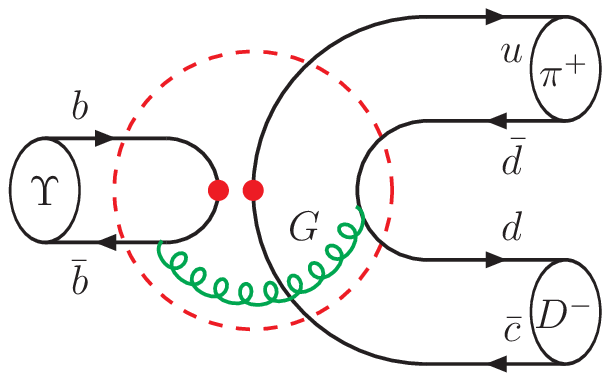} \\
  {(a) \hspace{0.21\textwidth} (b)
       \hspace{0.21\textwidth} (c)
       \hspace{0.21\textwidth} (d)}
  \caption{The Feynman diagram for the ${\Upsilon}(1S)$ ${\to}$
  $D^{-}{\pi}^{+}$ decay with the pQCD approach, where (a,b) are
  factorizable diagrams, and (c,d) are nonfactorizable diagrams.
  The dots denote appropriate interactions, and the dashed circles
  denote quark scattering amplitudes.}
  \label{feynman-pqcd}
  \end{figure}

  Recently, many QCD-inspired phenomenological models,
  such as the QCD factorization (QCDF) approach \cite{prl83.1914,
  npb591.313,npb606.245,plb488.46,plb509.263,prd64.014036}
  based on the collinear approximation, and the perturbative QCD
  (pQCD) approach \cite{prl74.4388,plb348.597,prd52.3958,
  prd63.074006,prd63.054008,prd63.074009,plb555.197}
  based on $k_{T}$ factorization, have been successfully applied
  to exclusive nonleptonic $B$ meson decay processes.
  Using these models, HMEs have a simple structure.
  They are generally expressed as the
  convolution of scattering sub-amplitudes arising from hard
  gluon exchanges among quarks and the wave functions (WFs)
  reflecting the nonperturbative contributions.
  The scattering sub-amplitudes are in principle calculable order
  by order with the perturbative theory.
  WFs are universal and process-independent, and could be
  obtained by nonperturbative methods or from data. So the
  theoretical calculation of HMEs becomes reasonably practical.
  The $W^{\pm}$ exchange topology of Fig. \ref{feynman-sm}
  corresponds to annihilation topologies (see Fig. \ref{feynman-pqcd})
  within the effective theory of Eq.(\ref{eq:hamilton}).
  Another two unknown parameters or more will be introduced to deal
  with the endpoint divergences of the annihilation amplitudes
  with the QCDF approach \cite{prd65.074001,prd65.094025,
  prd68.054003,npb675.333,npb774.64,prd90.054019,prd91.074026,
  plb740.56,plb743.444}.
  While the transverse momentum effects and Sudakov factors are
  considered to settle the endpoint contributions of quark
  scattering amplitudes and hadronic WFs with the pQCD approach
  \cite{prl74.4388,plb348.597,prd52.3958,prd63.074006,
  prd63.054008,prd63.074009,plb555.197}.
  In this paper, we will investigate the ${\Upsilon}(1S)$
  ${\to}$ $DP$ decays with the pQCD approach.
  The master pQCD formula for decay amplitudes could be factorized
  into three parts : the hard contributions above the scale of
  ${\mu}$ incorporated into the Wilson coefficients $C_{i}$,
  the perturbatively calculable quark scattering amplitudes
  ${\cal H}$ near the scale of ${\mu}$, and the long-distribution
  contributions below the scale of ${\mu}$ incorporated into
  hadronic WFs ${\Phi}$.
   \begin{eqnarray}
  {\cal A}_{i} &=& {\int} dx_{1}\,dx_{2}\,dx_{3}\,
   db_{1}\,db_{2}\,db_{3}\,C_{i}(t_{i})\,
  {\cal H}_{i}(x_{1},x_{2},x_{3},b_{1},b_{2},b_{3})
   \nonumber \\ & & \quad
  {\Phi}_{\Upsilon}(x_{1},b_{1})\,e^{-S_{\Upsilon}}\,
  {\Phi}_{D}(x_{2},b_{2})\,e^{-S_{D}}\,
  {\Phi}_{P}(x_{3},b_{3})\,e^{-S_{P}}
   \label{eq:pqcd-hme},
   \end{eqnarray}
  where $x_{i}$ is the longitudinal momentum fraction of the valence
  quark, $b_{i}$ is the conjugate variable of the transverse momentum,
  and $e^{-S_{i}}$ is the Sudakov factor.

  With the convention of Refs. \cite{plb752.322,ijmpa31.1650061,
  jhep0605.004,prd65.014007,prd78.014018}, the relevant mesonic WFs
  and distribution amplitudes (DAs) are defined as follows.
   \begin{eqnarray}
  {\langle}\,0\,{\vert}\,\bar{b}_{\alpha}(0)b_{\beta}(z)\,{\vert}
  {\Upsilon}(p_{1},{\epsilon}_{\parallel})\,{\rangle}\,
  =\, \frac{f_{\Upsilon}}{4}\,{\int}d^{4}k_{1}\,
   e^{-i\,k_{1}{\cdot}z}\, \big\{
   \!\!\not{\epsilon}_{\Upsilon}^{\parallel}
   \big[ m_{\Upsilon}\,{\phi}_{\Upsilon}^{v}
   -\!\not{p}_{1}\, {\phi}_{\Upsilon}^{t} \big]
   \big\}_{{\beta}{\alpha}}
   \label{eq:wf-def-upsilon},
   \end{eqnarray}
   \begin{equation}
  {\langle}\,\overline{D}(p_{2})\,{\vert}\,\bar{q}_{\alpha}(z)
  c_{\beta}(0)\, {\vert}\,0\,{\rangle}\, =\,
  -\frac{i\,f_{D}}{4}\,{\int}d^{4}k_{2}\,e^{+i\,k_{2}{\cdot}z}\,
   \big\{ {\gamma}_{5}\, \big( \!\!\not{p}_{2}+m_{D} \big)\,
  {\phi}_{D} \big\}_{{\beta}{\alpha}}
   \label{eq:wf-def-D-meson},
   \end{equation}
    \begin{eqnarray} & &
   {\langle}\,P(p_{3})\,{\vert}\, \bar{q}_{\alpha}(z)\,
    u_{\beta}(0)\, {\vert}\,0\,{\rangle}
    \nonumber \\ &=&
   -\frac{i\,f_{P}}{4}\, {\int}d^{4}k_{3}\,e^{+i\,k_{3}{\cdot}z}\,
    \big\{ {\gamma}_{5} \big[ \!\not{p}_{3}\,{\phi}_{P}^{a}
   +{\mu}_{P}\,{\phi}_{P}^{p}
   -{\mu}_{P}\,\big( \!\not{n}_{-}\!\!\not{n}_{+}-1\big)\,
    {\phi}_{P}^{t} \big] \big\}_{{\beta}{\alpha}}
    \label{eq:wf-def-pseudoscalar},
    \end{eqnarray}
  where $f_{\Upsilon}$, $f_{D}$ and $f_{P}$ are decay constants.
  ${\mu}_{P}$ $=$ $1.6{\pm}0.2$ GeV \cite{jhep0605.004} is the chiral
  mass. $n_{+}$ $=$ $(1,0,0)$ and $n_{-}$ $=$ $(0,1,0)$ are the
  light cone vectors, and satisfy the relations of $n_{\pm}^{2}$
  $=$ $0$ and $n_{+}{\cdot}n_{-}$ $=$ $1$.
  In the rest frame of the ${\Upsilon}(1S)$ meson,
  the kinematic variables are defined as follows.
   \begin{equation}
    p_{\Upsilon}\, =\, p_{1}\, =\, \frac{m_{\Upsilon}}{\sqrt{2}}
   \big(1,1,0\big)
   \label{kine-upsilon},
   \end{equation}
   \begin{equation}
    p_{D}\, =\, p_{2}\, =\, \frac{m_{\Upsilon}}{\sqrt{2}}
   \big(1,r_{D}^{2},0\big)
   \label{kine-D-meson},
   \end{equation}
   \begin{equation}
   p_{P}\, =\, p_{3}\, =\, \frac{m_{\Upsilon}}{\sqrt{2}}
   \big(0,1-r_{D}^{2},0\big)
   \label{kine-pion},
   \end{equation}
   \begin{equation}
    k_{1}\, =\, 
   \frac{m_{\Upsilon}}{\sqrt{2}}\big(x_{1},x_{1},\vec{k}_{1T}\big)
   \label{kine-b-quark},
   \end{equation}
   \begin{equation}
   k_{2}\, =\, 
   \frac{m_{\Upsilon}}{\sqrt{2}}\big(x_{2},0,\vec{k}_{2T}\big)
   \label{kine-D-quark},
   \end{equation}
   \begin{equation}
   k_{3}\, =\, 
   \frac{m_{\Upsilon}}{\sqrt{2}}\big(0,x_{3}\,
   (1-r_{D}^{2}),\vec{k}_{3T}\big)
   \label{kine-pion-quark},
   \end{equation}
   \begin{equation}
  {\epsilon}_{\Upsilon}^{\parallel}\, =\, \frac{1}{ \sqrt{2} }
   \big(1,-1,0\big)
   \label{kine-polarization-vector},
   \end{equation}
  where $k_{i}$, $x_{i}$ and $\vec{k}_{iT}$ are respectively
  the momentum, longitudinal momentum fraction and
  transverse momentum, as shown in Fig. \ref{feynman-pqcd}(a).
  The mass ratio $r_{D}$ $=$ ${\displaystyle \frac{m_{D}}{m_{\Upsilon}} }$.
  ${\epsilon}_{\Upsilon}^{\parallel}$ is the longitudinal
  polarization vector. The explicit DA expressions
  \cite{plb752.322,ijmpa31.1650061,
  jhep0605.004,prd65.014007,prd78.014018} are as follows.
   \begin{equation}
  {\phi}_{\Upsilon}^{v}(x) =  A\, x\,\bar{x}\,
  {\exp}\big\{ -\frac{m_{b}^{2}}{8\,{\beta}_{1}^{2}\,x\,\bar{x}} \big\}
   \label{da-bbv},
   \end{equation}
   \begin{equation}
  {\phi}_{\Upsilon}^{t}(x) = B\, {\xi}^{2}\,
  {\exp}\big\{ -\frac{m_{b}^{2}}{8\,{\beta}_{1}^{2}\,x\,\bar{x}} \big\}
   \label{da-bbt},
   \end{equation}
   \begin{equation}
  {\phi}_{D}(x) = C\, x\,\bar{x}\,
  {\exp}\big\{ -\frac{1}{8\,{\beta}_{2}^{2}}\,
   \big( \frac{m_{q}^{2}}{x}+\frac{m_{c}^{2}}{\bar{x}} \big) \big\}
   \label{da-cqa},
   \end{equation}
   \begin{equation}
  {\phi}_{D}(x,b) = 6\,x\,\bar{x}\,\big\{1-C_{D}\,
  {\xi} \big\}\, {\exp}\big\{ -\frac{1}{2}\,
  {\omega}_{D}^{2}\,b^{2} \big\}
   \label{wave-cq},
   \end{equation}
    \begin{equation}
   {\phi}_{P}^{a}(x)\, =\, 6\,x\,\bar{x}\,\big\{
    1+a_{1}^{P}\,C_{1}^{3/2}({\xi})
     +a_{2}^{P}\,C_{2}^{3/2}({\xi})\big\}
    \label{wf-pi-twsit-2},
    \end{equation}
    \begin{eqnarray}
   {\phi}_{P}^{p}(x) &=& 1+3\,{\rho}_{+}^{P}
   -9\,{\rho}_{-}^{P}\,a_{1}^{P}
   +18\,{\rho}_{+}^{P}\,a_{2}^{P}
    \nonumber \\ &+&
    \frac{3}{2}\,({\rho}_{+}^{P}+{\rho}_{-}^{P})\,
    (1-3\,a_{1}^{P}+6\,a_{2}^{P})\,{\ln}(x)
    \nonumber \\ &+&
    \frac{3}{2}\,({\rho}_{+}^{P}-{\rho}_{-}^{P})\,
    (1+3\,a_{1}^{P}+6\,a_{2}^{P})\,{\ln}(\bar{x})
    \nonumber \\ &-&
    (\frac{3}{2}\,{\rho}_{-}^{P}
    -\frac{27}{2}\,{\rho}_{+}^{P}\,a_{1}^{P}
    +27\,{\rho}_{-}^{P}\,a_{2}^{P})\,C_{1}^{1/2}(\xi)
    \nonumber \\ &+&
    ( 30\,{\eta}_{P}-3\,{\rho}_{-}^{P}\,a_{1}^{P}
    +15\,{\rho}_{+}^{P}\,a_{2}^{P})\,C_{2}^{1/2}(\xi)
    \label{wf-pi-twsit-3-p},
    \end{eqnarray}
    \begin{eqnarray}
   {\phi}_{P}^{t}(x) &=&
    \frac{3}{2}\,({\rho}_{-}^{P}-3\,{\rho}_{+}^{P}\,a_{1}^{P}
    +6\,{\rho}_{-}^{P}\,a_{2}^{P})
    \nonumber \\ &-&
    C_{1}^{1/2}(\xi)\big\{
    1+3\,{\rho}_{+}^{P}-12\,{\rho}_{-}^{P}\,a_{1}^{P}
   +24\,{\rho}_{+}^{P}\,a_{2}^{P}
    \nonumber \\ & & \quad +
    \frac{3}{2}\,({\rho}_{+}^{P}+{\rho}_{-}^{P})\,
    (1-3\,a_{1}^{P}+6\,a_{2}^{P})\,{\ln}(x)
    \nonumber \\ & & \quad +
    \frac{3}{2}\,({\rho}_{+}^{P}-{\rho}_{-}^{P})\,
    (1+3\,a_{1}^{P}+6\,a_{2}^{P})\, {\ln}(\bar{x}) \big\}
    \nonumber \\ &-&
    3\,(3\,{\rho}_{+}^{P}\,a_{1}^{P}
    -\frac{15}{2}\,{\rho}_{-}^{P}\,a_{2}^{P})\,C_{2}^{1/2}(\xi)
    \label{wf-pi-twsit-3-t},
    \end{eqnarray}
   where $\bar{x}$ $=$ $1$ $-$ $x$ and ${\xi}$ $=$ $x$ $-$
   $\bar{x}$ $=$ $2\,x$ $-$ $1$.
   ${\beta}_{1}$ $=$ $m_{b}\,{\alpha}_{s}(m_{b})$
   and ${\beta}_{2}$ $=$ $m_{D}\,{\alpha}_{s}(m_{D})$
   are the shape parameters of DAs for the ${\Upsilon}(1S)$ and $D$
   mesons. $a_{i}^{P}$ and $C_{n}^{m}(\xi)$ are the Gegenbauer
   moment and Gegenbauer polynomials. The other shape parameters
   of pseudoscalar DAs are ${\rho}_{+}^{P}$ $=$
   ${\displaystyle  \frac{m_{P}^{2}}{{\mu}_{P}^{2}} }$,
   ${\rho}_{-}^{K}$ ${\simeq}$ ${\displaystyle \frac{m_{s}}{{\mu}_{K}} }$,
   ${\rho}_{-}^{\pi}$ $=$ $0$,
   and ${\eta}_{P}$ $=$ ${\displaystyle \frac{f_{3P}}{f_{P}\,{\mu}_{P}} }$.
   The parameters $A$, $B$ and $C$ in Eq.(\ref{da-bbv}),
   Eq.(\ref{da-bbt}) and Eq.(\ref{da-cqa}) can be
   determined by the normalization conditions.
    \begin{equation}
   {\int}_{0}^{1}dx\, {\phi}_{\Upsilon}^{v,t}(x)\, =\, 1
    \label{da-bbv-normalization-conditions},
    \end{equation}
    \begin{equation}
   {\int}_{0}^{1}dx\, {\phi}_{D_{q}}(x)\, =\,1
    \label{wf-D-normalization-conditions}.
    \end{equation}

  It should be pointed out there are many models for DAs
  of the $D$ mesons, for example, Eq.(30) of Ref.\cite{prd78.014018}.
  In this paper, we take the typical models of Eq.(\ref{da-cqa})
  and Eq.(\ref{wave-cq}) for examples to illustrate the model
  dependence of the results.
  For the scenario I of Eq.(\ref{da-cqa}), the mass of light quark
  is $m_{u,d}$ $=$ $310$ MeV and $m_{s}$ $=$ $510$ MeV \cite{book2014}.
  For the scenario II of  Eq.(\ref{wave-cq}), the shape parameters
  $C_{D}$ $=$ $0.5$ and ${\omega}_{D}$ $=$ $0.1$ GeV for the
  $D_{u,d}$ meson and $C_{D}$ $=$ $0.4$ and ${\omega}_{D}$
  $=$ $0.2$ GeV for the $D_{s}$ meson \cite{prd78.014018}.

  The lowest order Feynman diagrams for the ${\Upsilon}(1S)$
  ${\to}$ $D{\pi}$ decay with the pQCD approach are shown
  in Fig. \ref{feynman-pqcd}.
  After a series of calculation with the pQCD formula of
  Eq.(\ref{eq:pqcd-hme}), the expressions of the decay
  amplitude and branching ratio are written as follows.
   \begin{eqnarray}
  {\cal A}({\Upsilon}{\to}DP) &=& F\,
   \frac{G_{F}}{\sqrt{2}}\, V_{ub} V_{cb}^{\ast}\,
   \big\{ \big( a_{2}-a_{3}-a_{5}+\frac{1}{2}a_{7}
  +\frac{1}{2}a_{9} \big)\,\big( {\cal A}_{a}^{LL}
  +{\cal A}_{b}^{LL} \big)
   \nonumber \\ & & \hspace{-0.1\textwidth}
  +\big( C_{1}-C_{4}+\frac{1}{2}C_{10} \big)\,
   \big( {\cal A}_{c}^{LL}+{\cal A}_{d}^{LL} \big)
  -\big(C_{6}-\frac{1}{2}C_{8} \big)\,
   \big( {\cal A}_{c}^{LR}+{\cal A}_{d}^{LR} \big) \big\}
   \label{eq:decay-amplitude},
   \end{eqnarray}
   \begin{equation}
  {\cal B}r \,=\,
   \frac{p_{\rm cm}}{24\,{\pi}\,m_{\Upsilon}^{2}\,
  {\Gamma}_{\Upsilon}}\,
  {\vert} {\cal A}({\Upsilon}{\to}DP) {\vert}^{2}
   \label{eq:branching-ratio},
   \end{equation}
  where the factor $F$ $=$ ${\displaystyle \frac{1}{\sqrt{2}} }$
  for $P$ $=$ ${\pi}^{0}$, and $F$ $=$ $+1$ for $P$ $=$
  ${\pi}^{+}$ and $K^{+}$. $p_{\rm cm}$ is the
  center-of-mass momentum of final states in the rest
  frame of the ${\Upsilon}(1S)$ meson.
  The building blocks of amplitudes
  ${\cal A}_{i}^{j}$ are listed in Appendix \ref{blocks}.
  With the input parameters in Table \ref{tab:input-parameter},
  the numerical results of the branching ratios for the
  ${\Upsilon}(1S)$ ${\to}$ $DP$ decays are summarized
  in Table \ref{tab:br}.

   \begin{table}[ht]
   \caption{The values of the input parameters, where their
   central values are regarded as the default inputs
   unless otherwise specified. The numbers in parentheses
   are errors.}
   \label{tab:input-parameter}
   \begin{ruledtabular}
   \begin{tabular}{ccc}
    \multicolumn{3}{c}{mass and decay constants of the particles \cite{pdg2020} } \\ \hline
    $m_{{\pi}^{0}}$ $=$ $134.98$ MeV,
  & $m_{D^{0}}$ $=$ $1864.84(5)$ MeV,
  & $f_{{\pi}}$ $=$ $130.2(1.2)$ MeV, \\
    $m_{{\pi}^{\pm}}$ $=$ $139.57$ MeV,
  & $m_{D^{\pm}}$ $=$ $1869.5(4)$ MeV
  & $f_{K}$ $=$ $155.7(3)$ MeV, \\
    $m_{K^{\pm}}$ $=$ $493.68$ MeV,
  & $m_{D_{s}^{\pm}}$ $=$ $1969.0(1.4)$ MeV,
  & $f_{3{\pi}}$ $=$ $0.45(15){\times}10^{-2}\,{\rm GeV}^{2}$ \cite{jhep0605.004}, \\
    $m_{b}$ $=$ $4.78(6)$ GeV,
  & $f_{D}$ $=$ $212.6(7)$ MeV,
  & $f_{3K}$ $=$ $0.45(15){\times}10^{-2}\,{\rm GeV}^{2}$ \cite{jhep0605.004}, \\
    $m_{c}$ $=$ $1.67(7)$ GeV,
  & $f_{D_{s}}$ $=$ $249.9(5)$ MeV,
  & $f_{{\Upsilon}(1S)}$ $=$ $676.4(10.7)$ MeV \cite{prd92.074028}, \\ \hline
    \multicolumn{3}{c}{Gegenbauer moments at the scale of ${\mu}$
    $=$ 1 GeV \cite{jhep0605.004}} \\ \hline
    \multicolumn{3}{c}{ $a_{1}^{\pi}$ $=$ $0$, \qquad
    $a_{2}^{\pi}$ $=$ $0.25(15)$, \qquad
    $a_{1}^{K}$ $=$ $0.06(3)$, \qquad
    $a_{2}^{K}$ $=$ $0.25(15)$ }
  \end{tabular}
  \end{ruledtabular}
  \end{table}
   \begin{table}[h]
   \caption{Branching ratios for the ${\Upsilon}(1S)$ ${\to}$
   $DP$ decays in the unit of $10^{-18}$.
   The uncertainties come from the shape parameters of all
   participating hadronic DAs,
   including $m_{b}$, $m_{c}$, ${\mu}_{P}$ and $a_{2}^{P}$
   for scenario I and
   $m_{b}$, $C_{D}{\pm}0.2$, ${\omega}_{D}{\pm}0.04$ GeV,
   ${\mu}_{P}$ and $a_{2}^{P}$ for scenario II, respectively.}
   \label{tab:br}
   \begin{ruledtabular}
   \begin{tabular}{cccc}
   mode & $D^{-}{\pi}^{+}$
        & $\overline{D}^{0}{\pi}^{0}$
        & $D_{s}^{-}K^{+}$ \\ \hline
  scenario I
        & $ 1.16^{+ 0.24}_{- 0.19}$
        & $ 0.58^{+ 0.12}_{- 0.09}$
        & $ 1.78^{+ 0.38}_{- 0.29}$ \\
  scenario II
        & $ 0.77^{+ 0.27}_{- 0.19}$
        & $ 0.39^{+ 0.13}_{- 0.10}$
        & $ 1.29^{+ 0.55}_{- 0.33}$
  \end{tabular}
  \end{ruledtabular}
  \end{table}

  (1)
  For each specific process, the branching ratios of
  scenario I is larger than those of scenario II.
  The branching ratios are sensitive to the DA models for the
  $D$ mesons.

  (2)
  Because of the relations among decay constants, {\em i.e.},
  $f_{D_{s}}$ $>$ $f_{D}$ and $f_{K}$ $>$ $f_{\pi}$,
  there is a clear hierarchical pattern among branching ratios.
   \begin{equation}
  {\cal B}r({\Upsilon}(1S){\to}D_{s}K)\, >\,
  {\cal B}r({\Upsilon}(1S){\to}D_{d}{\pi})\, >\,
  {\cal B}r({\Upsilon}(1S){\to}D_{u}{\pi})
   \label{bf-hierarchical-relation}.
   \end{equation}
  In addition, there is a relation, ${\cal B}r({\Upsilon}(1S){\to}D_{d}{\pi})$
  ${\approx}$ $2\,{\cal B}r({\Upsilon}(1S){\to}D_{u}{\pi})$,
  because of the quark compositions of electrically neutral pion,
  {\em i.e.}, ${\vert}{\pi}^{0}{\rangle}$ $=$
  ${\displaystyle \frac{{\vert}u\bar{u}{\rangle}-
   {\vert}d\bar{d}{\rangle}}{\sqrt{2}} }$.

  (3)
  Hadronic DAs are the essential parameters of the amplitudes.
  The theoretical uncertainties from the shape parameters
  of hadronic DAs are given in Table \ref{tab:br}.
  Besides, another 7\% and 3\% uncertainties can come from the
  CKM factor ${\vert}V_{ub}\,V_{cb}^{\ast}{\vert}$ and
  decay constants, respectively.

  (4)
  There are many possible reasons for the small branching ratios.
  (a)
  Almost all of the decay width of the ${\Upsilon}(1S)$ meson
  come from the strong and electromagnetic interactions.
  The ${\Upsilon}(1S)$ weak decays are strongly suppressed by
  the Fermi coupling constant $G_{F}$ ${\sim}$ $10^{-5}$,
  compared with the couplings of
  ${\alpha}_{s}$ ${\sim}$ $10^{-1}$
  and ${\alpha}_{em}$ ${\sim}$ $10^{-2}$.
  (b)
  The annihilation amplitudes are generally
  power suppressed based on the power counting rule
  in the heavy quark limit \cite{npb591.313}.
  (c)
  These processes are highly suppressed by the CKM factor of
  ${\vert}V_{ub}\,V_{cb}^{\ast}{\vert}$ ${\sim}$ $10^{-4}$.
  (d)
  According to the conservation law of angular momentum,
  only the $P$-wave contribution allows in
  the decay amplitudes.
  (e)
  These decays are suppressed by color due to the $W$ exchange
  between quarks of different final states.

  (5)
  The branching ratios for the ${\Upsilon}(1S)$ ${\to}$ $DP$ decays
  within SM are at
  the order of $10^{-18}$, which are too small to be
  measurable in the near future.
  Of course, it is possible that some extraordinary effects from
  NP may significantly enhance these branching ratios, and produce
  an observable phenomena.

  In summary, considering the developmental opportunities and
  important challenges at the high-luminosity dedicated
  heavy-flavor factories in the future,
  the exclusive two-body nonleptonic ${\Upsilon}(1S)$ decays
  through the weak interactions into final states including
  only one charmed meson, ${\Upsilon}(1S)$ ${\to}$ $DP$, are
  studied for the first time with the pQCD approach.
  Our results show that (1) the ${\Upsilon}(1S)$ ${\to}$
  $D_{s}K$ decay has relatively large occurrence probability
  among the concerned processes; (2) the branching ratios for
  the ${\Upsilon}(1S)$ ${\to}$ $DP$ decay are tiny,
  ${\cal O}(10^{-18})$, and impossible to measure at Belle-II
  and LHCb during the next decades.
  One experimental signal of the ${\Upsilon}(1S)$
  ${\to}$ $D_{s}K$ and/or $D{\pi}$ decays will be an obvious
  deviation from the SM prediction and an omen of NP.

  \section*{Acknowledgments}
  The work is supported by the National Natural Science Foundation
  of China (Grant Nos. 11705047, 11981240403, U1632109 and 11547014).

  \begin{appendix}
  \section{Building blocks of decay amplitudes}
  \label{blocks}
  For the sake of convenience in writing, some shorthands are used.
  There should always be a Sudakov factor corresponding to each WF
  with the pQCD approach.
  So the shorthands are ${\phi}_{\Upsilon}^{v,t}$ $=$
  ${\phi}_{\Upsilon}^{v,t}(x_{1})\,e^{-S_{\Upsilon}}$,
  ${\phi}_{D}$ $=$ ${\phi}_{D}(x_{2})\,e^{-S_{D}}$,
  ${\phi}_{P}^{a}$ $=$ ${\phi}_{P}^{a}(x_{3})\,e^{-S_{P}}$ and
  ${\phi}_{P}^{p,t}$ $=$ ${\displaystyle \frac{{\mu}_{P}}{m_{\Upsilon}}\,
   {\phi}_{P}^{p,t}(x_{3})\,e^{-S_{P}} }$.
   \begin{equation}
   a_{i}\, =\, \bigg\{ \begin{array}{lll}
   \displaystyle C_{i}+\frac{1}{N_{c}}C_{i+1},
   & \quad & \text{for odd }i; \\
   \displaystyle C_{i}+\frac{1}{N_{c}}C_{i-1},
   & \quad & \text{for even }i.
   \end{array} 
   \label{wilson-ai}
   \end{equation}

  According to the pQCD formula of Eq.(\ref{eq:pqcd-hme}),
  the amplitude building block ${\cal A}_{i}^{j}$ should be
  a function of the Wilson coefficients $C_{i}$. That is to say,
  the expression $C_{k}\,{\cal A}_{i}^{j}$ in
  Eq.(\ref{eq:decay-amplitude}) should actually be
  ${\cal A}_{i}^{j}[C_{k}]$.
  As to the amplitude building block ${\cal A}_{i}^{j}$,
  the subscript $i$ corresponds to the indices of
  Fig.\ref{feynman-pqcd}, and the superscript
  $j$ refers to the two possible Dirac structures
  ${\Gamma}_{1}{\otimes}{\Gamma}_{2}$ of the operator
  $(\bar{q}_{1}{\Gamma}_{1}q_{2})(\bar{q}_{3}{\Gamma}_{2}q_{4})$,
  namely $j$ $=$ $LL$ for ${\gamma}^{\mu}(1-{\gamma}_{5})
  {\otimes}{\gamma}_{\mu}(1-{\gamma}_{5})$ and $j$ $=$ $LR$
  for ${\gamma}^{\mu}(1-{\gamma}_{5}){\otimes}{\gamma}_{\mu}(1+{\gamma}_{5})$.
  The expressions of $\mathcal{A}_{i}^{j}$ are written as follows.
   \begin{equation}
  {\cal C}\, =\, m_{\Upsilon}^{4}\,f_{\Upsilon}\,f_{D}\,f_{P}
   \frac{{\pi}\,C_{F}}{N_{c}}
   \label{eq:coefficient-block},
   \end{equation}
   \begin{eqnarray}
  {\cal A}_{a}^{LL}  &=& {\cal C} {\int}_{0}^{1}dx_{2}\,dx_{3}
  {\int}_{0}^{\infty}db_{2}\,db_{3}\, {\alpha}_{s}(t_{a})\,
  H_{ab}({\alpha}_{g},{\beta}_{a},b_{2},b_{3})\,C_{i}(t_{a})
   \nonumber \\ & &
   S_{t}(x_{2})\, {\phi}_{D}\,\big\{ {\phi}_{P}^{a}\,(1-r_{D}^{2})\,x_{2}
  +2\,r_{D}\,{\phi}_{P}^{p}\,  \big[ x_{2}-(1-r_{D}^{2}) \big] \big\}
   \label{eq:amp-fig-a},
   \end{eqnarray}
   \begin{eqnarray}
  {\cal A}_{b}^{LL}  &=& {\cal C} {\int}_{0}^{1}dx_{2}\,dx_{3}
  {\int}_{0}^{\infty}db_{2}\,db_{3}\, {\alpha}_{s}(t_{b})\,
  H_{ab}({\alpha}_{g},{\beta}_{b},b_{3},b_{2})\,C_{i}(t_{b})
   \nonumber \\ & &
   S_{t}(x_{3})\, {\phi}_{D}\,\big\{
  {\phi}_{P}^{a}\,(1-r_{D}^{2})\,
   \big[ x_{3}\,(1-r_{D}^{2})-r_{D}^{2} \big]
   \nonumber \\ & &
  +r_{D}\,(1-r_{D}^{2})\,(x_{3}-\bar{x}_{3})
   \big[ {\phi}_{P}^{p}+{\phi}_{P}^{t} \big]
   +2\,r_{D}\,{\phi}_{P}^{t} \big\}
   \label{eq:amp-fig-b},
   \end{eqnarray}
   \begin{eqnarray}
  {\cal A}_{c}^{LL}  &=& \frac{{\cal C}}{N_{c}}
  {\int}_{0}^{1}dx_{1}\,dx_{2}\,dx_{3}
  {\int}_{0}^{\infty}db_{1}\,db_{2}\, {\alpha}_{s}(t_{c})\,
  H_{cd}({\alpha}_{g},{\beta}_{c},b_{1},b_{2})\,C_{i}(t_{c})
   \nonumber \\ & &
  {\phi}_{D}\,\big\{ {\phi}_{\Upsilon}^{v}\, \big[
  {\phi}_{P}^{a}\,(1-r_{D}^{2})\,\big( x_{1}\,(1+r_{D}^{2})
  -x_{2}\,r_{D}^{2}-x_{3}\,(1-r_{D}^{2}) \big)
   \nonumber \\ & &
  +r_{D}\,{\phi}_{P}^{p}\, \big( x_{2}-x_{3}\,(1-r_{D}^{2}) \big)
  +r_{D}\,{\phi}_{P}^{t}\, \big( 2\,x_{1}-x_{2}-x_{3}\,
   (1-r_{D}^{2}) \big) \big]
   \nonumber \\ & &
  -\frac{1}{2}\,{\phi}_{\Upsilon}^{t} \big[
  {\phi}_{\pi}^{a}\,(1-r_{D}^{2})+4\,r_{D}\,{\phi}_{P}^{t}
   \big] \big\}_{b_{2}=b_{3}}
   \label{eq:amp-fig-c-LL},
   \end{eqnarray}
   \begin{eqnarray}
  {\cal A}_{c}^{LR}  &=& \frac{{\cal C}}{N_{c}}
  {\int}_{0}^{1}dx_{1}\,dx_{2}\,dx_{3}
  {\int}_{0}^{\infty}db_{1}\,db_{2}\, {\alpha}_{s}(t_{c})\,
  H_{cd}({\alpha}_{g},{\beta}_{c},b_{1},b_{2})\,C_{i}(t_{c})
   \nonumber \\ & &
  {\phi}_{D}\,\big\{ {\phi}_{\Upsilon}^{v}\, \big[
  {\phi}_{P}^{a}\,(1-r_{D}^{2})^{2}\,(x_{2}-x_{1})
  +r_{D}\,{\phi}_{P}^{p}\, \big( x_{2}-x_{3}\,(1-r_{D}^{2}) \big)
   \nonumber \\ & & 
  -r_{D}\,{\phi}_{P}^{t}\, \big( 2\,x_{1}-x_{2}-x_{3}\,(1-r_{D}^{2})
   \big) \big] +\frac{1}{2}\,{\phi}_{\Upsilon}^{t}\,
  {\phi}_{P}^{a}\,(1-r_{D}^{2}) \big\}_{b_{2}=b_{3}}
   \label{eq:amp-fig-c-LR},
   \end{eqnarray}
   \begin{equation}
  {\cal A}_{d}^{LL}\, =-\, {\cal A}_{c}^{LR}
  (x_{1}{\to}\bar{x}_{1},\,t_{c}{\to}t_{d},\,
  {\beta}_{c}{\to}{\beta}_{d})
   \label{eq:amp-fig-d-LL},
   \end{equation}
   \begin{equation}
  {\cal A}_{d}^{LR}\, =-\, {\cal A}_{c}^{LL}
  (x_{1}{\to}\bar{x}_{1},\,t_{c}{\to}t_{d},\,
  {\beta}_{c}{\to}{\beta}_{d})
   \label{eq:amp-fig-d-LR},
   \end{equation}
   \begin{eqnarray} & &
   H_{ab}({\alpha},{\beta},b_{i},b_{j})\, =\,
  -\frac{{\pi}^{2}}{4}\,b_{i}\,b_{j}\,
   \big\{ J_{0}(b_{j}\sqrt{{\alpha}})
      +i\,Y_{0}(b_{j}\sqrt{{\alpha}}) \big\}
   \nonumber \\ & & \qquad
   \big\{ {\theta}(b_{i}-b_{j})
   \big[ J_{0}(b_{i}\sqrt{{\beta}})
     +i\,Y_{0}(b_{i}\sqrt{{\beta}}) \big]
      J_{0}(b_{j}\sqrt{{\beta}})
   + (b_{i}{\leftrightarrow}b_{j}) \big\}
   \label{eq:amp-denominator-ab},
   \end{eqnarray}
   \begin{eqnarray} & &
   H_{cd}({\alpha},{\beta},b_{1},b_{2})\, =\,
   b_{1}\,b_{2}\,
   \big\{ \frac{i{\pi}}{2}\,{\theta}({\beta})
   \big[ J_{0}(b_{1}\sqrt{{\beta}})
     +i\,Y_{0}(b_{1}\sqrt{{\beta}})
   \big]+{\theta}(-{\beta})
   K_{0}(b_{1}\sqrt{-{\beta}}) \big\}
   \nonumber \\ & & \qquad
   \frac{i{\pi}}{2}\, \big\{ {\theta}(b_{1}-b_{2})
   \big[ J_{0}(b_{1}\sqrt{{\alpha}})
     +i\,Y_{0}(b_{1}\sqrt{{\alpha}}) \big]
      J_{0}(b_{2}\sqrt{{\alpha}})
   + (b_{1}{\leftrightarrow}b_{2}) \big\}
   \label{eq:amp-denominator-cd},
   \end{eqnarray}
   \begin{equation}
   S_{\Upsilon}\, =\, s(x_{1},p_{1}^{+},1/b_{1})
   +2\,{\int}_{1/b_{1}}^{t} \frac{d{\mu}}{{\mu}}{\gamma}_{q}
   \label{eq:sudakov-upsilon},
   \end{equation}
   \begin{equation}
   S_{D}\, =\, s(x_{2},p_{2}^{+},1/b_{2})+s(\bar{x}_{2},p_{2}^{+},1/b_{2})
   +2\,{\int}_{1/b_{2}}^{t} \frac{d{\mu}}{{\mu}}{\gamma}_{q}
   \label{eq:sudakov-D-meson},
   \end{equation}
   \begin{equation}
   S_{P}\, =\, s(x_{3},p_{3}^{-},1/b_{3})+s(\bar{x}_{3},p_{3}^{-},1/b_{3})
   +2\,{\int}_{1/b_{3}}^{t} \frac{d{\mu}}{{\mu}}{\gamma}_{q}
   \label{eq:sudakov-pseudoscalar},
   \end{equation}
   \begin{equation}
  {\alpha}_{g}\, =\, m_{\Upsilon}^{2}\,(1-r_{D}^{2})\,x_{2}\,x_{3}
   \label{eq:amp-gluon},
   \end{equation}
   \begin{equation}
  {\beta}_{a}\, =\, m_{\Upsilon}^{2}\,(1-r_{D}^{2})\,x_{2}
   \label{eq:amp-quark-fig-a},
   \end{equation}
   \begin{equation}
  {\beta}_{b}\, =\, m_{\Upsilon}^{2}\,(1-r_{D}^{2})\,x_{3}
   \label{eq:amp-quark-fig-b},
   \end{equation}
   \begin{equation}
  {\beta}_{c}\, =\, {\alpha}_{g}-
   m_{\Upsilon}^{2}\,(1-r_{D}^{2})\,x_{1}\,x_{3}
  -m_{\Upsilon}^{2}\,x_{1}\,x_{2}
   \label{eq:amp-quark-fig-c},
   \end{equation}
   \begin{equation}
  {\beta}_{d}\, =\, {\alpha}_{g}-
   m_{\Upsilon}^{2}\,(1-r_{D}^{2})\,\bar{x}_{1}\,x_{3}
  -m_{\Upsilon}^{2}\,\bar{x}_{1}\,x_{2}
   \label{eq:amp-quark-fig-d},
   \end{equation}
   \begin{equation}
   t_{a,b}\, =\, {\max}(\sqrt{{\beta}_{a,b}},
                  1/b_{2},1/b_{3})
   \label{eq:amp-scale-fig-a-b},
   \end{equation}
   \begin{equation}
   t_{c,d}\, =\, {\max}(\sqrt{{\alpha}_{g}},
        \sqrt{{\vert}{\beta}_{c,d}{\vert}},
                  1/b_{1},1/b_{2})
   \label{eq:amp-scale-fig-c-d},
   \end{equation}
  where $I_{0}$, $J_{0}$, $K_{0}$ and $Y_{0}$ are Bessel functions.
  The expression of $s(x,Q,b)$ can be found in Ref.\cite{prd52.3958}.
  ${\gamma}_{q}$ $=$ ${\displaystyle -\frac{{\alpha}_{s}}{\pi} }$
  is the quark anomalous dimension.
  \end{appendix}

  
  \end{document}